\documentclass{desyproc}

\begin{document}
\title{Search for a Scalar Axion--like Particle at 34 GHz}

\author{{\slshape
P. L. Slocum$^1$,  O. K. Baker$^1$, J. L. Hirshfield$^{1,2}$, Y.
Jiang$^1$, G. Kazakevitch$^3$, S. Kazakov$^1$, M. A. LaPointe$^1$,
A. T. Malagon$^1$, A. J. Martin$^1$, S. Shchelkunov$^1$, A.
Szymkowiak$^{1,4}$}\\[1ex]
$^1$Department of Physics, Yale University, PO Box 208120, New
Haven, CT USA 06520\\
$^2$Omega-P, Inc., 291 Whitney Ave., Suite 401, New Haven, CT
06511\\
$^3$Muons, Inc., 552 N. Batavia Avenue, Batavia, IL 60510\\
$^4$Department of Astronomy, Yale University, PO Box 208101, New
Haven CT  06520}

\contribID{familyname\_firstname}

\desyproc{DESY-PROC-2012-04}
\acronym{Patras 2012} 
\doi  

\maketitle

\begin{abstract}
Light axion--like particles ({\small ALP}s) that couple to two
photons are allowed in a number of proposed extensions to the
Standard Model of elementary particles.  Of particular interest from
a theoretical and observational standpoint is the energy regime near
0.1 meV.  We present results from a pilot experiment to search for a
signal from a 0.14 meV scalar {\small ALP} by way of its coupling to
two photons. Using a copper resonant cavity cooled to four degrees
Kelvin while immersed in a seven Tesla magnetic field, and coupled
to a low noise cryogenic amplifier and room temperature receiver, we
exclude an {\small ALP}--driven excess of 34~GHz photons with
g~$>$10$^{-8}$/GeV with 5$\sigma$ confidence. We discuss the
ramifications of this initial measurement as well as planned
modifications to the experiment for increased sensitivity.
\end{abstract}

\section{Introduction}

Several theories of particle physics as well as cosmology predict
the existence of at least one sub--eV scalar, that is, spin-zero,
boson \cite{ahlers1, jaeckelringwald1, holdom1, foot1, jrr1}.
Correspondingly, many theories of physics beyond the Standard Model
(SM) can accommodate scalars with very small masses and feeble
couplings to SM fields \cite{choi, wf1}.  An intriguing possibility
in astrophysics and cosmology is that these weakly interacting,
sub--eV particles ({\small WISP}s) may constitute at least some
component of the cold dark matter in the universe \cite{sikivie1,
vg1, Arias:2012mb}. It has been shown that these arguments apply for
both pseudoscalar, namely axion~\cite{sikivie1, vg1}, and
scalar~\cite{Arias:2012mb} {\small WISP}s, e.g. axion--like
particles ({\small ALPs}) that couple to two photons.

The current experimental programs that explore the parameter space
of weakly interacting, light, spin-zero bosons include sensitive
searches that make use of resonant cavities \cite{sikivie2, bradley,
cjr1, jr1}. Scalar {\small ALP}s may be detected by stimulation of
the conversion to photons in a magnetic field, similar to the way
that axion conversion may take place \cite{ymce1}.  This conversion
can occur in a suitable cavity that is properly tuned to the mass of
the {\small ALP}.

While it should be noted that scalar couplings to two photons are
strongly excluded by fifth force experiments~\cite{adelberger}, it
is possible that a low energy form factor could relax these
constraints in some models~\cite{adelberger}. The Lagrangian that
describes the coupling of a light scalar particle to two photons in
the presence of an external magnetic field takes the form:

\begin{equation}\nonumber
{\rm L}_{S\gamma\gamma} ={1 \over 4}
g_{S\gamma\gamma}S^0F_{\mu\nu}F^{\mu\nu} \ = \
-g_{S\gamma\gamma}S^0B \cdot B_{ext}
\end{equation}

\noindent where $F_{\mu\nu}$ is the electromagnetic field strength
tensor, $g_{S\gamma\gamma}$ is the strength of the coupling between
the scalar particle $S^0$ and two photons, $B$ is the magnetic field
of the incident photon, and $B_{ext}$ is the external magnetic
field. From this Lagrangian, the equations of motion that result may
be used to write the power $P_{S\gamma}$ that results from axion to
photon conversion on resonance in the microwave cavity
as~\cite{asztalos}
\begin{equation}\label{eq:power}
P_{S\gamma} = g_{S\gamma\gamma}^2 V B_{ext}^2 \rho_a C_{lmn} Q.
\end{equation}
\noindent Here, $V$ is the cavity volume, $C_{lmn}$ is the form
factor associated with the cavity mode, $Q$ is the quality factor of
the cavity, and $\rho_a$ is the assumed scalar {\small ALP} density.
The results presented here are the first from a search for
primordial scalar particles coupling to a strong external magnetic
field using resonant cavities in the mass range of 0.14 meV
corresponding to 34 GHz frequency.

\section{Experiment}\label{sectionExperiment}
The apparatus~\cite{ymce1,slocum2009,martin} consists of a tunable,
34~GHz resonant cavity (Q$\sim$10$^4$) made from oxygen-free copper
and a high electron mobility transistor ({\small HEMT}) cryogenic
amplifier~\cite{weinreb} located at the bottom of a cold gas
cryostat that is oriented vertically and cooled to approximately
4~K.  The cavity and cryogenic amplifier are coupled by
approximately 10~cm of {\small WR28} waveguide. The cryostat rests
inside the vertical bore of a 7~T cryomagnet.  The temperature
inside the cryostat is monitored at multiple locations with
cryogenic thin film resistance temperature sensors.

The resonant cavity has one critically coupled and one weakly
coupled port, each connected to {\small WR28} waveguide.  The
critically coupled waveguide terminates at the input to the {\small
HEMT} amplifier, after which both waveguides feed out of the
cryostat. The air in the cavity is pumped out through the weakly
coupled waveguide.  The Q of the cavity is measured prior to each
data run using a network analyzer connected to both waveguides.

The cavity is tuned with an adjustable plunger that is vacuum tight
at 4~K.  A fiberglass G-10 rod is threaded through the top of the
cryostat and fastened to the tuning plunger with a horizontal lever.
When the G-10 rod is turned, the plunger moves vertically. The range
of tuning in the cavity is $\pm$5~mm which corresponds to
approximately $\pm$0.8 GHz.

The signal from the cryogenic amplifier terminates outside the
cryostat at a waveguide--to--coaxial adapter and triple heterodyne
microwave receiver~\cite{ymce1}. After the receiver the voltages
from the in--phase and quadrature components are digitized for the
complex Fast Fourier Transform ({\small FFT}) and further analysis
offline.  The sensitivity of the experiment is limited by the system
noise temperature $T_{sys}$ according to the Dicke radiometer
equation~\cite{dicke}.
\begin{equation}\nonumber
\sigma_T=\frac{T_{sys}}{\sqrt{\Delta \nu \tau}}.
\end{equation}

\section{Measurements and Results}

\begin{figure}
\includegraphics[width=3.0in]{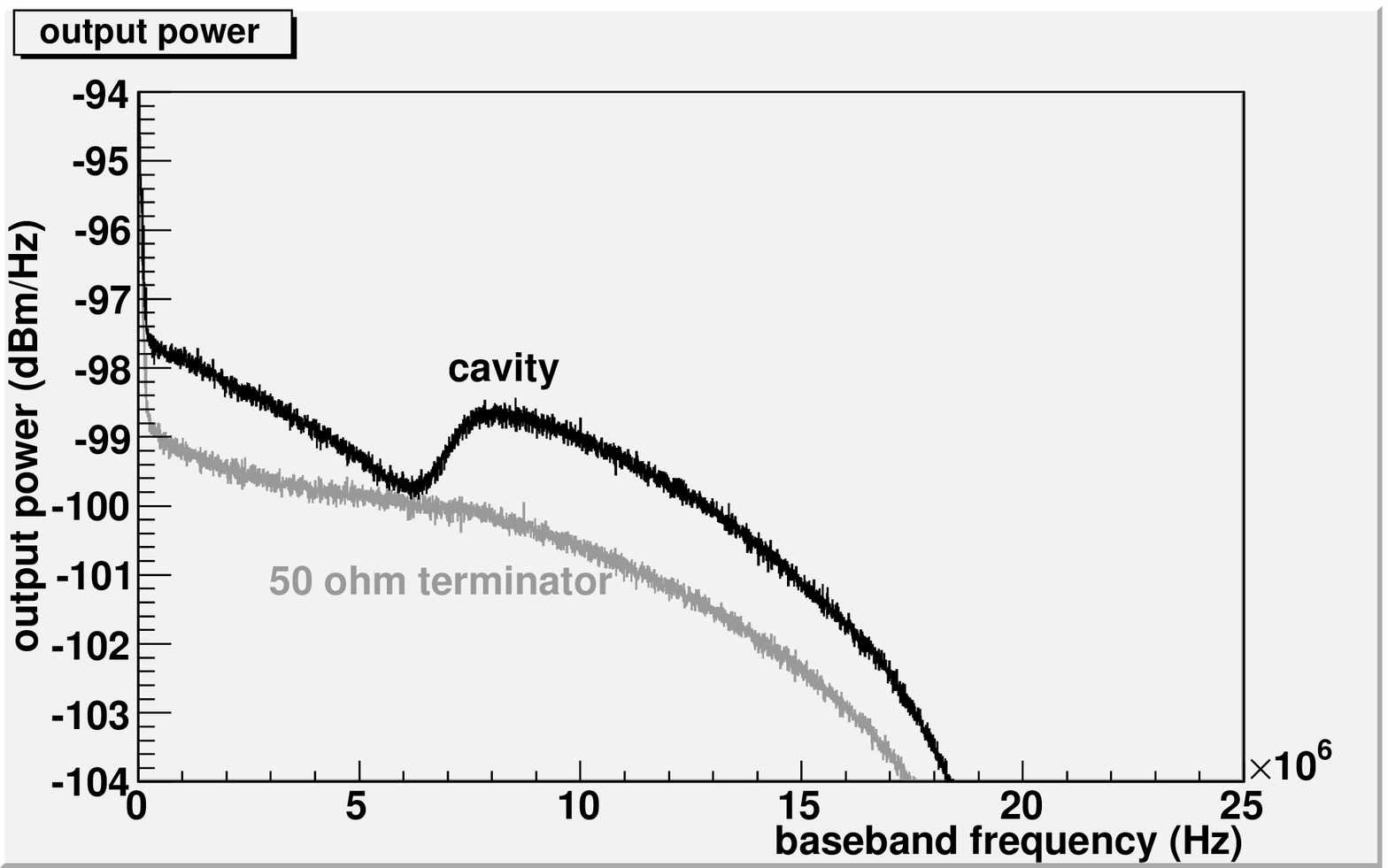}
\includegraphics[width=3.0in]{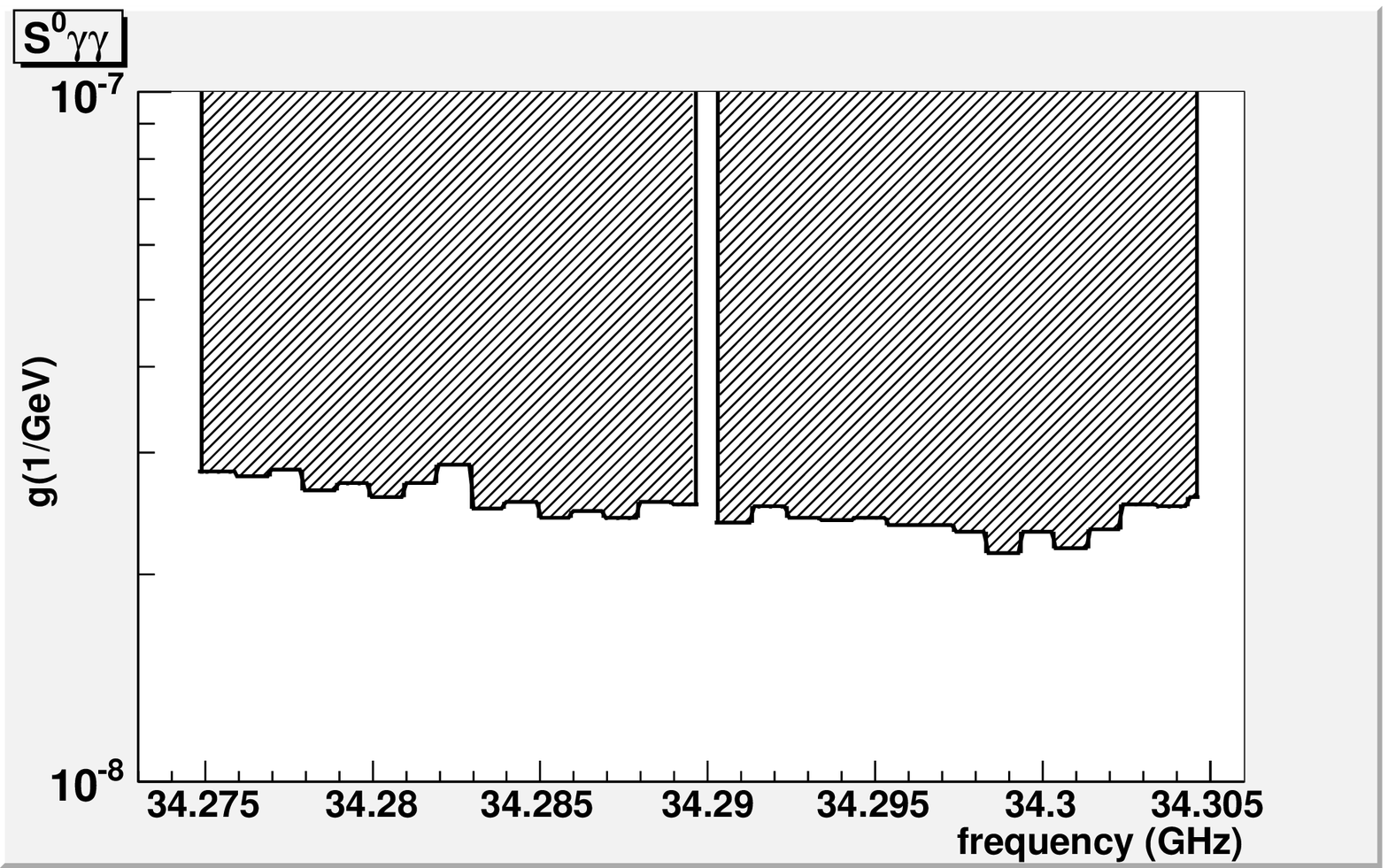}
\caption{Left panel:  Two power spectra measured with the first HEMT
cryoamplifier held at T~=~6~K.  The spectrum marked {\em cavity}
shows a raw measurement of the power at the output of the receiver
chain with the cold resonant cavity tuned to 34.409~GHz (6-7~MHz on
the plot). The trace labeled {\em terminator} shows the power
measured at the output of the receiver with a cold 40~GHz 50-ohm
terminator at the input to the first HEMT. In both measurements the
shape of the spectrum is driven by the last bandpass filter in the
electronics chain. Right panel: Exclusion limit on $S^0\gamma\gamma$
coupling from data similar to those shown in the left panel.  In
this initial result the measurements are centered at 34.29~GHz
(0.142 meV) and span across the width of the last bandpass filter
(30~MHz). The gap in the center of the plot corresponds to
frequencies at or near 0~Hz in the baseband and have been excluded
from this analysis.\label{fig:results}}
\end{figure}

Typical power spectra measured with the cryogenic amplifier held at
6 K are shown in the left panel of Figure~\ref{fig:results}.  The
resonance of the cavity appears as a dip in the spectrum.  This dip
is qualitatively consistent with the combination of two effects: A
frequency-dependent reflection of noise power coming from the input
of the HEMT, and a change in the gain of the HEMT as the impedance
of the source decreases (e.g.~\cite{wedge,meys}). The width of the
dip is approximately 3-4 MHz which is in accordance with the
cavity's Q of 10$^4$.

The expected power from couplings between a scalar {\small ALP} and
2 photons is given in Equation~\ref{eq:power}~\cite{asztalos}. The
cavity form factor $C_{lmn}$, adapted from \cite{asztalos} for the
case of a scalar {\small ALP}, is
\begin{equation}\nonumber
C_{lmn} \equiv \frac{\left | \int_V {d^3x {\bf B} \cdot {\bf \hat
B}_{\text{ext}}}\right |^2}{V \int_V{d^3x \frac{1}{\mu}\left|{\bf
B}\right|^2}},
\end{equation}
where ${\bf B}e^{i\omega t}$ is the oscillating magnetic field in
the cavity, ${\bf \hat B}_{\text{ext}}$ is the static magnetic
field, and $\mu$ is the magnetic permeability.  For the
configuration used in these measurements $C_{lmn}$=10$^{-6}$ is
lower than would be desirable, but the cavity geometry was dictated
by the constraints of a different experiment~\cite{slocum2009}.
Using a limited set of measurements similar to those in the left
panel of Figure~\ref{fig:results} and assuming a primordial {\small
ALP} density $\rho_a$ of 10$^{13}$/cm$^3$~\cite{gates}, the right
panel shows that we exclude couplings with g$>$10$^{-8}$/GeV between
0.14 meV scalar particles and two photons with 5$\sigma$ confidence.

Although the sensitivity of the present measurement does not exceed
model-dependent limits on g$_{S\gamma\gamma}$ set in previous
searches for solar {\small ALP}s~\cite{zioutas}, by astrophysical
observations~\cite{raffelt}, and by fifth force
experiments~\cite{adelberger}, it is the first glimpse into this
energy regime with a technique that has mass resolution. It is also
the first direct search for {\small ALP}s as cold dark matter at 0.1
meV. Immediate plans for the experiment include a cavity in a
transverse magnetic mode which will allow coupling to pseudoscalar
{\small ALP}s and will be 3 orders of magnitude more sensitive than
the present measurement. A wider mass range will also be searched by
tuning the receiver. In conclusion although this measurement is
primarily a first step toward the goal of a more sensitive
experiment, it is still an unprecedented, narrow band test of
$S^0\gamma\gamma$ coupling limits that are otherwise
model-dependent.

\section{Acknowledgments}
This work was supported by the Office of Naval Research award
N00014--09--1--0481.

\section{Bibliography}

\begin{footnotesize}

\end{footnotesize}


\end{document}